\documentclass[12pt]{article}

\input epsf
\usepackage{epsfig,graphicx,epsf,hyperref}
\usepackage{amsmath,amsfonts,amssymb,amsbsy}
\hypersetup{colorlinks=true,linkcolor=blue,citecolor=blue,urlcolor=blue}

\def\i{\mbox{i}}
\def\\i{\mbox{\scriptsize{i}}}
\def\d{\mbox{d}}

\font\eulerRM = eurm10 scaled \magstep1
\def\R{\mbox{\eulerRM\char'032}}

\title{High-energy Coulomb scattering of spatially extended particles
}
\author{M. L. Nekrasov\\
{\small\it 
Institute for High Energy Physics, NRC ``Kurchatov
Institute'',}  \vspace*{-4\baselineskip}\\
{\small\it Protvino 142281, Russia} }
\date{}

\begin{document}
\maketitle

\begin{abstract}
We analyze pure Coulomb high-energy elastic scattering of charged particles (hadrons or nuclei), discarding their strong interactions. We distinguish three scattering modes, determined by the magnitude of the momentum transfer, in which particles behave as point-like, structureless extended, and structured composite objects. The results are compared in the potential and QFT approaches of the eikonal model. In the case of proton Coulomb scattering at the LHC the difference between these two approaches is significant. This indicates the unsuitability of the potential approach. However, in the case of Coulomb scattering of heavy nuclei, the leading one is the optical approximation, which formally reproduces the prescription of the potential approach. 
\end{abstract}

\section{Introduction}\label{sec1}

At low momentum transfer the Coulomb and strong interactions make compete contributions to the elastic scattering of charged particles. In this regard, the determination of Coulomb contributions and then the description of Coulomb-nuclear interference (CNI) is an essential part of the study of strong interactions of hadrons in the small-transfer region. Meanwhile, despite the impressive history of the issue \cite{Bethe}-\cite{Nekrasov2024} the problem of the Coulomb contributions has not yet been completely solved. In particular, there is no consensus on how to account for the spatial sizes of charged particles, such as protons. Similar problems arise when describing the Coulomb scattering of heavy nuclei with large electric charges.

At high energies, the generally accepted method for describing Coulomb scattering is based on the eikonal model, which ensures the summation of Coulomb contributions in scattering at small angles. However, in different versions of this model, the eikonal phase is defined differently. Most studies use the potential version, see e.g.~\cite{Islam1967}-\cite{Petrov2024} and references therein, in which the eikonal phase, as well as the Born contribution, are determined through the classical potential interaction of electric charges. In the quantum field theory (QFT) version of the model, the eikonal phase and the Born contribution are determined by the simplest diagram with the one-photon exchange in the $t$-channel, see e.g.~\cite{Cheng-Wu,Levy-Sucher}. In the case of point-like particles the mentioned difference is ultimately unimportant. However, in the case of spatially extended particles it leads to different prescriptions for taking into account the electromagnetic form factors.
Namely, in the potential approach, particle form factors directly enter the eikonal phase, while in the QFT form factors arise when averaging the full amplitude operator between the wave functions of in- and out-states, and therefore appear as factors in the full amplitude. 

In fact, both above approaches are based on the idea of particles as extended structureless objects. However, the mode of multiple Coulomb scattering on their structural components is also possible. We mean the scattering on charged partons in the case of hadrons and on charged protons in the case of scattering of nuclei. In both cases, each act of elementary scattering must contribute to the Born amplitude. Such a scattering mode was described by Glauber \cite{Glauber1959} in relation to elastic scattering due to strong interactions of a fast point-like particle off a nucleus. Subsequently, many attempts were made to generalize this approach to the case of nucleus-nucleus scattering, each time using the short-range property of nuclear forces, see e.g.~\cite{Czyz}-\cite{Shabelski}. However, the generalization that is important for heavy nuclei to the case of scattering due to long-range Coulomb forces has not yet been consistently obtained. Accordingly, there is no generalization to the case of Coulomb scattering of hadrons on their charged partons.

In this article, we explore the above issues and determine which scattering mode actually occurs and under what conditions. The key point will be the idea that the scattering mode is determined by the magnitude of the momentum transfer. In particular, at the very small transfers ($q^2 \lesssim 10^{-2} \,\mbox{GeV}^{2}$ in the case of protons) the wavelength of the exchanged photons is of the order or exceeds the transverse sizes of colliding particles. In this case they scatter as structureless objects. At higher transfers, the scattering can occur on the charged components of the particles. In the latter case, the scattering must be described by the Glauber theory. We consistently construct its generalization to the case of Coulomb scattering of composite objects and show that the so-called optical approximation is indeed the leading one if the scattering objects contain a large number of charged components. 

In the next section we discuss Coulomb scattering  of point-like particles. The generalization to the case of extended particles in different scattering modes is given in Sect.~\ref{sec3}. There we also carry out numerical simulations with different methods of accounting form factors. In the final section we discuss the results.

\section{Coulomb scattering of point-like particles \label{sec2}}

Following the notation of \cite{Cahn,Nekrasov2024}, we define the amplitude of elastic scattering of point-like particles at high energies in the eikonal approach as follows:
\begin{equation}\label{A1}
F(s,q^2) = 
\frac{s}{4\pi \i} \int \d^2 {\bf b} \;\; 
e^{\\i {\bf q b}} \left[ e^{2\\i\delta(s,{\bf b})} - 1 \right].
\end{equation}
Here $s$ is the square of the energy of colliding particles in the center-of-mass system, ${\bf q}$ is the momentum transfer, $q^2 = {\bf q}^2 = -t$, ${\bf b}$ is the impact parameter, $\delta(s,{\bf b})$ is the phase shift. From here on, two-dimensional vectors in the plane of impact parameter are  denoted in bold without an arrow at the top. The differential cross section is determined by the amplitude (\ref{A1}) as follows,
\begin{equation}\label{A2}
\frac{\d \sigma}{\d t} = \frac{\pi}{s p^2_{cm}} |F|^2\,,
\end{equation} 
where $p^2_{cm} = s/4$. Expression (\ref{A1}) is a consequence of the well-known formula for the expansion of the amplitude in partial waves at large momenta and small scattering angles. Up to this limitation, formula (\ref{A1}) is model-free.

A specific model is determined by the phase shift $\delta(s,{\bf b})$. In the potential approach, it is determined by the interaction potential of particles \cite{Glauber1959,Moliere1947},
\begin{equation}\label{A3}
\delta(s,{\bf b}) = - \, \frac{1}{2} \, \int_{-\infty}^{+\infty}
\d z \; V(s,\vec{\bf r}) \,.
\end{equation}
Hereinafter $\vec{\bf r} = ({\bf b},z)$, where $z$ is the longitudinal coordinate. The potential $V(s,\vec{\bf r})$ simultaneously defines the Born amplitude,
\begin{equation}\label{A4}
F_{\mbox{\scriptsize Born}}(s,{\bf q}^2) =  - \frac{s}{4\pi}\, \int \d^3 \vec{\bf r} \;
e^{\\i \vec{\bf q} \vec{\bf r}} \; V(s,\vec{\bf r}){\bigl|_{q_z = 0 }\bigr.}\,.
\end{equation}
So $\delta(s,{\bf b})$ may be written as
\begin{equation}\label{A5}
\delta(s,{\bf b}) = 
\frac{2\pi}{s} \int \frac{\d^2 {\bf q}}{(2\pi)^2} \; 
e^{-\\i {\bf q b}} F_{\mbox{\scriptsize Born}}(s,{\bf q}^2) \,.
\end{equation}

It should be noted here that the derivation of formula (\ref{A3}) is based on the non-relativistic Schr{\"{o}dinger equation. This implies that the total energy of the system is divided into potential and kinetic parts, and the latter part is determined (in the c.m.s.) by the non-relativistic formula $E_{kin} = p^2_{cm}/2m$, where $m$ is the reduced mass. We emphasize that the latter condition is essential for deriving formula (\ref{A3}). For this reason, extending the scope of application of this formula to the case of relativistic scattering is, at least, a strong model assumption. Fortunately, the final formulas for the amplitude turn out to be (formally) relativistically invariant. {\sl A posteriori}, this justifies the use of this approach, but only until another, more suitable approach appears. We will return to discuss this issue below.

At the same time, in some studies formula (\ref{A3}) is initially postulated, implying that it is valid in relativistic theory. In this case $V(s,\vec{\bf r})$ is referred to as the optical potential. Unfortunately, it is very difficult to give it a constructive definition. In particular, solving the inverse problem, it is possible to determine only the integral of it, as given in (\ref{A3}). Nevertheless, it can be argued \cite{Arnold1967} that the optical potential differs from other previously defined potentials, such as the quasi-potential of Logunov and Tavkhelidze, the mass shell potential of Chew and Frautschi, etc. 

For the above reason, in the QFT approach the phase shift is defined directly using formula (\ref{A5}) with the appropriate definition of $F_{\mbox{\scriptsize Born}}$. For instance, if $F_{\mbox{\scriptsize Born}}$ is the simplest contribution due to the exchange in the $t$-channel by some intermediate particle acting as a quantum of interaction, then formula (\ref{A1}) describes the result of summing a generalized ladder of similar exchanges in the approximation of large $s$ and restricted $t$ \cite{Cheng-Wu,Levy-Sucher}. Of course, in the general case certain radiative corrections should be added to $F_{\mbox{\scriptsize Born}}$ defined in this way, but in the case of quantum electrodynamics they are minor and do not make significant contributions to the asymptotic formula~(\ref{A1}).

So, next we consider the case of electrodynamics. Then in the potential approach the potential for point-like particles with identical charges, such as protons, is $V^{C}(\vec{\bf r}) = \alpha /|\vec{\bf r}|$, where $\alpha$ is the fine structure constant. Correspondingly, the Born amplitude according to (\ref{A4}) is 
\begin{equation}\label{A6}
F^C_{\mbox{\scriptsize Born}} = 
- \frac{\alpha s}{{\bf q}^2 + \lambda^2}\,.
\end{equation} 
Here $\lambda$ is a parameter introduced to regularize IR divergences, which should eventually be directed to zero. In the QFT approach, the Coulomb Born amplitude is described by the same formula (\ref{A6}). So, in the case of point-like particles, the formulations of the eikonal model for Coulomb scattering in the potential and QFT approaches completely coincide.

Substituting (\ref{A6}) into (\ref{A5}), we get the phase shift
\begin{equation}\label{A7}
\delta^{C}({\bf b}) = 
\frac{1}{2\pi} \int \d^2 {\bf q} \; 
e^{-\\i {\bf q b}} \frac{-\alpha}{q^2 + \lambda^2} =
- \alpha \mbox{K}_0 (b \lambda)\,,
\end{equation}
where $\mbox{K}_0 (z)$ is the McDonald function of the zero order. Substituting (\ref{A7}) into formula (\ref{A1}), we obtain the following result:
\begin{equation}\label{A8}
F^{C}(s,q^2){\Bigl|_{\lambda \to 0 }\Bigr.} = 
- \, \frac{s \alpha}{q^2} \, e^{\\i \alpha\ln (\lambda^2/q^2)} 
e^{\\i \Phi(\alpha)} + \i s \pi \delta({\bf q})\,.
\end{equation}
Here the last term on the r.h.s.~compensates for the delta function that arises in the first term at $\alpha \to 0$, see details in \cite{Nekrasov2024}. So at $\alpha \to 0$~amplitude (\ref{A8}) becomes zero in accordance with (\ref{A1}). The $\Phi(\alpha) = 2 \alpha\gamma - \i
\ln \left\{\Gamma(1+\i\alpha) / \Gamma(1-\i \alpha)\right\}$ in (\ref{A8}) is a real value. Correspondingly, $[\i \alpha\ln (\lambda^2/q^2) + \i \Phi(\alpha)]$ determines a phase that accumulates IR-divergent contributions in accordance with general theorems on the elimination of IR divergences. The emergence of the phase is a consequence of the summation of the generalized ladder of Born contributions. For ${\bf q} \not= 0$ the phase does not contribute to the cross section and can be omitted along with the $\delta$-function.

\section{Spatially extended particles
\label{sec3}}

The formalism considered above is quite applicable for describing scattering of extended charged particles in the case of extremely small values of transferred momenta, when the wavelengths of the exchanged photons significantly exceed the transverse sizes of colliding particles. As the transfer increases and the wavelengths of photons become comparable to the particle sizes, their sizes must be taken into account. In particular, if the wavelength of a photon exceeds but is comparable to the sizes of the particles, then the time of emission and absorption of the photon exceeds the time of interaction between internal parts of the particles. As a result, the scattering occurs immediately on all of their charged parts, and the photon exchange occurs between the volumes of particles without reference to specific points of emission and absorption. (In essence, these points are of a probabilistic nature.) Colliding particles in this case look like structureless objects equipped with a continuous charge distribution density. With a further increase in $q^2$, when the wavelength of photons becomes smaller than the distance between centers of charged components of the particles, Coulomb scattering should increasingly occur on these components.\footnote{An exception is the case of strong correlation between the particle components. In hadrons, this regime is predicted in some models at ultra-high energies \cite{Nekrasov2020,Nekrasov2021,Nekrasov2022}. In this case the region of the momentum transfer in which hadrons are characterized by a continuous charge distribution density increases. \label{foot2}} In the case of nuclei such charged components are protons, and in the case of hadrons they are charged partons.

Both the above modes can be analyzed in the potential and QFT approaches, and the results, generally speaking, may not coincide. In case of discrepancy, its cause should be understood. Below we begin to analyze this issue, starting with the scattering mode in which particles behave as structureless extended objects. In the case of protons, this mode occurs at $q^2 \lesssim 10^{-2} \,\mbox{GeV}^{2}$, which roughly corresponds to distances of two or more of their radii. (In the same range CNI effect is significant at the LHC energies \cite{Kaspar2011,TOTEM8,TOTEM13}.) In the case of nuclei, for example Pb-Pb Coulomb scattering, a similar estimate is $q^2 \lesssim 2 \times 10^{-4}\,\mbox{GeV}^{2}$. The cases with higher momentum transfers, we consider in further subsections.

\subsection{Particles with continuous charge distribution
 \label{sec3.1}}

In the potential approach, a generally accepted way to generalize eikonal model to the case of scattering of extended particles is to introduce form factors in the eikonal phase. Actually, such a generalization should follow from the appropriate modification of the definition of  potential in formula (\ref{A3}). Below we show this in the case when scattering particles are structureless objects characterized by a continuous charge distribution. In doing so, we will pay  special attention to issues of relativistic invariance. In particular, we will take into account the phenomenon of potential retardation in reference frames where both particles are moving.

So, let particles $a$ and $b$, each with electric charge $\sqrt{\alpha}$, have charge distribution densities $\R_{\kappa}(\vec{\bf r}_{\kappa})$, $\kappa = a,b$. In this case $\d q_{\kappa} = \sqrt{\alpha}\,\d\vec{\bf r}_{\kappa}\R_{\kappa}(\vec{\bf r}_{\kappa})$ is the elementary electric charge located in the spatial volume $\d\vec{\bf r}_{\kappa}$, and in any reference frame the charge distribution density $\R_{\kappa}(\vec{\bf r}_{\kappa})$ is normalized by the condition
\begin{equation}\label{A9}
\int \d\vec{\bf r}_{\kappa} \R_{\kappa}(\vec{\bf r}_{\kappa}) = 1\,.
\end{equation} 

First, we consider the rest frame of particle $a$. In this case, the elementary charge $\d q_{a}$ forms a static electric potential $\d\phi_a(\vec{\bf r}) = \d q_a/|\vec{\bf r}|$, where $\vec{\bf r}$ specifies a point in space relative to the location $\d q_{a}$. Now let particle $b$ move in the direction of particle $a$ along the $z$ axis with impact parameter ${\bf b}$. If an elementary charge $\d q_{b}$ appears in the point $\vec{\bf r}$, then the potential energy of interaction of these elementary charges is $\delta V^{C}_{ab}(\vec{\bf r}) = \d q_a \d q_b /|\vec{\bf r}|$. Hence, the potential energy of interaction of particles $a$ and $b$ in this reference frame is 
\begin{equation}\label{A10}
V^{C}_{ab}(\vec{\bf r}) = 
\int \d \vec{\bf r}_a \R_a(\vec{\bf r}_a) \,
\d \vec{\bf r}_b \R_b(\vec{\bf r}_b) \,
\frac{\alpha}{|\vec{\bf r} - \vec{\bf r}_a + \vec{\bf r}_b|} \,,
\end{equation} 
where $\vec{\bf r}$ determines the position of the center of particle $b$ relative to the center of particle $a$, $\vec{\bf r}_a$ and $\vec{\bf r}_b$ are the radius-vectors of elementary charges relative to the centers of the corresponding particles. Next, we put $\vec{\bf r} = ({\bf b},z)$, $\vec{\bf r}_{a} = ({\bf s}_a,z_a)$, $\vec{\bf r}_{b} = ({\bf s}_b,z_b)$, substitute (\ref{A10}) into (\ref{A3}) assuming that the latter formula is valid in the rest frame of particle $a$, and integrate first over $\d z$. Then we obtain a formula for the Coulomb shift in the case of spatially extended particles,
\begin{equation}\label{A11}
\delta^{C}_{ab}({\bf b}) =  
\int \d {\bf s}_a \d {\bf s}_b \, 
\rho_a({\bf s}_a) \rho_b({\bf s}_b) \,
\delta^{C}({\bf b} - {\bf s}_a + {\bf s}_b)\,.
\end{equation}
Here $\rho_{a}({\bf s}_{a})$ and $\rho_{b}({\bf s}_{b})$ are the result of integration $\R_{a}(\vec{\bf r}_{a})$ and $\R_{b}(\vec{\bf r}_{b})$ over $\d z_{a}$ and $\d z_b$, respectively, and $\delta^C$ is the elementary Coulomb shift, specified in (\ref{A7}). Passing to momentum space and taking into account (\ref{A5}) and (\ref{A6}), we get 
\begin{equation}\label{A12}
\delta^{C}_{ab}({\bf b}) =   
\frac{1}{2\pi} \! \int \d^2 {\bf q} \, 
e^{-\\i {\bf q b}}
\frac{-\alpha}{{\bf q}^2 + \lambda^2} \;
{\cal F}_{a}(q^2){\cal F}_{b}(q^2)\,,
\end{equation}  
where ${\cal F}_{a}(q^2)$ and ${\cal F}_{b}(q^2)$ are the Fourier of the transverse charge densities ${\rho}_{a}({\bf s})$ and ${\rho}_{b}({\bf s})$, i.e.~the particle form factors. Formula (\ref{A12}) is usually postulated in studies within the potential approach. 

The amplitude of Coulomb scattering at a given phase shift (\ref{A12}) is written as
\begin{equation}\label{A13}
F^{C}(s,q^2) = 
\frac{s}{4\pi \i} \int \d^2 {\bf b} \;\; 
e^{\\i {\bf q b}} \left[ e^{2\\i{\delta}^{C}_{ab}(b)} - 1 \right].
\end{equation}
The IR divergence in (\ref{A13}) is extracted into a phase. At ${\bf q} \not= 0$ the result is \cite{Petrov2018}
\begin{equation}\label{A14}
F^{C}(s,q^2) e^{-2\\i{\delta}^{C}_{ab}(0)} = 
\frac{s}{4\pi \i} \int \d^2 {\bf b} \;\; 
e^{\\i {\bf q b}} e^{2\\i\hat{\delta}^{C}_{ab}(b)} \,,
\end{equation}
where $\hat{\delta}^{C}_{ab}(b)$ is IR finite and at $\lambda = 0$ is equal to
\begin{equation}\label{A15}
\hat{\delta}^{C}_{ab}(b) =   
- \alpha \! \int_{0}^{\infty} \frac{\d q}{q} \, {\cal F}_{a}(q^2){\cal F}_{b}(q^2) \left[J_0(qb)-1\right]\,.
\end{equation}  
In this formula $J_0(z)$ is the Bessel function of the zero order. 

Now we note that amplitude (\ref{A14}) is relativistically invariant, but it is not clear why this happened, since expressions (\ref{A11}) and (\ref{A12}) were derived from a formula obtained in a relativistically non-invariant approach and applied in a specific reference frame.

To understand the reason, we derive above formulas in another reference frame, in which both particles are moving along $z$ axis. Since from the point of view of this reference frame our previous reference frame is moving, we must prime all non-invariant variables in formula (\ref{A10}), and our task now is to express all primed variables in variables of the new reference frame. In this regard, we first note that under Lorentz transformations the electric potential is transformed as the time component of a 4-vector. Therefore, since the vector potential in the rest frame of particle $a$ is equal to zero, in the new reference frame the energy of interaction of elementary charges is $\delta V^{C}_{ab}(\vec{\bf r}\,') = \gamma \, \d q_a \d q_b /|\vec{\bf r}\,'|$, where $\gamma = (1\!-\!\beta^2)^{-1/2}$ and $\beta$ is the velocity of particle $a$. So, in the new reference frame the interaction potential is equal~to
\begin{equation}\label{A16}
V^{C}_{ab} = 
\gamma\int \d \vec{\bf r}\,'_a \R'_a(\vec{\bf r}\,'_a) \,
\d \vec{\bf r}\,'_b \R'_b(\vec{\bf r}\,'_b) \,
\frac{\alpha}{\sqrt{({\bf b} - {\bf s}_a + {\bf s}_b)^2 + (z' - z'_a + z'_b)^2}} \,.
\end{equation} 
Here the r.h.s.~is written in terms of coordinates in the rest frame of particle $a$. We have to express them in terms of coordinates in the new reference frame. To do this, we synchronize both frames in such a way that at the initial moment of time the origins of the coordinates of both frames coincide with the center of particle $a$. Let, further, in the previous reference frame the longitudinal coordinate of the center of particle $b$ is $z'$ at time $t'$. Then the corresponding coordinate and time in the new reference frame are $z=\gamma(z' + \beta t')$ and $t=\gamma(t' + \beta z')$, and therefore $z' = \gamma (z-\beta t)$. In addition, since the elementary charge $\d q_a$ at the moment $t'$ occupies the position $z'_a$, we similarly have $t_a =\gamma(t' + \beta z'_a)$ and $z_a=\gamma(z'_a + \beta t')$, and therefore $z'_a = \gamma (z_a-\beta t_a)$. The situation with the coordinate $z'_b$ of the elementary charge $\d q_b$ is a little more complicated, since $z'_b = \tilde{z}'_b - z'$, where $\tilde{z}'_b$ is the position of $\d q_b$ relative to the center of particle $a$. With this in mind, similar calculations lead to $z'_b = \gamma (z_b-\beta t_b+\beta t)$, where $t_b =\gamma(t' + \beta \tilde{z}'_b)$. Substituting all this into (\ref{A16}) and taking into account $\d \vec{\bf r}\,'_{\kappa} \R'_{\kappa}(\vec{\bf r}\,'_{\kappa}) = \d \vec{\bf r}_{\kappa} \R_{\kappa}(\vec{\bf r}_{\kappa})$, we get
\begin{equation}\label{A17}
V^{C}_{ab} = 
\gamma\int \d \vec{\bf r}_a \R_a(\vec{\bf r}_a) \,
\d \vec{\bf r}_b \R_b(\vec{\bf r}_b) \,
\frac{\alpha}{\sqrt{({\bf b} - {\bf s}_a + {\bf s}_b)^2 + 
\gamma^2 (z - z_a + z_b + \beta t_a - \beta t_b)^2}} \,.
\end{equation} 

Let us again substitute (\ref{A17}) into formula (\ref{A3}), assuming that this formula is valid in the new reference frame. Then, as before, by first integrating over $\d z$, we arrive at (\ref{A11}) and (\ref{A12}), and then at the final formula (\ref{A14}).

So, the reason for the relativistic invariance of the Coulomb shift $\delta^{C}_{ab}({\bf b})$ obtained by formula (\ref{A3}) is the integration in this formula over the longitudinal coordinate.  However, this does not eliminate the problem of the non-relativistic genesis of formula (\ref{A3}), since its derivation is based on non-relativistic Schr{\"{o}dinger equation. Of course, there are other equations that use the concept of particle interaction potential, but we do not know a derivation of formula (\ref{A3}) based on them. In this regard, formulas (\ref{A11}) and (\ref{A12}) for the Coulomb shift need to be confirmed or refuted in a completely relativistic theory.

As such a theory, we consider quantum electrodynamics. The relevant studies were carried out in \cite{Nekrasov2024}. In this case the Born amplitude is defined in (\ref{A6}), and the total amplitude is determined by averaging the result of the eikonal summation of the generalized ladder of similar contributions with the wave functions of the initial and final states. The result is a simple modification of formula (\ref{A8}) by multiplying it by the product of form factors,
\begin{equation}\label{A18}
F^{C}(s,q^2){\Bigl|_{\lambda \to 0 }\Bigr.} = 
- \, \frac{s \alpha}{q^2} {\cal F}_{a}(q^2){\cal F}_{b}(q^2)\, e^{\\i \alpha\ln (\lambda^2/q^2)} 
e^{\\i \Phi(\alpha)} + \i s \pi \delta({\bf q})\,.
\end{equation}
So, at $q^2 \to 0$, since ${\cal F}_{\kappa}(q^2) \to 1$, the amplitude (\ref{A18}) turns into the amplitude of point-like particles.

\begin{figure*}
\begin{center}\hspace*{-0.06\textwidth}
\includegraphics[width=0.8\textwidth]{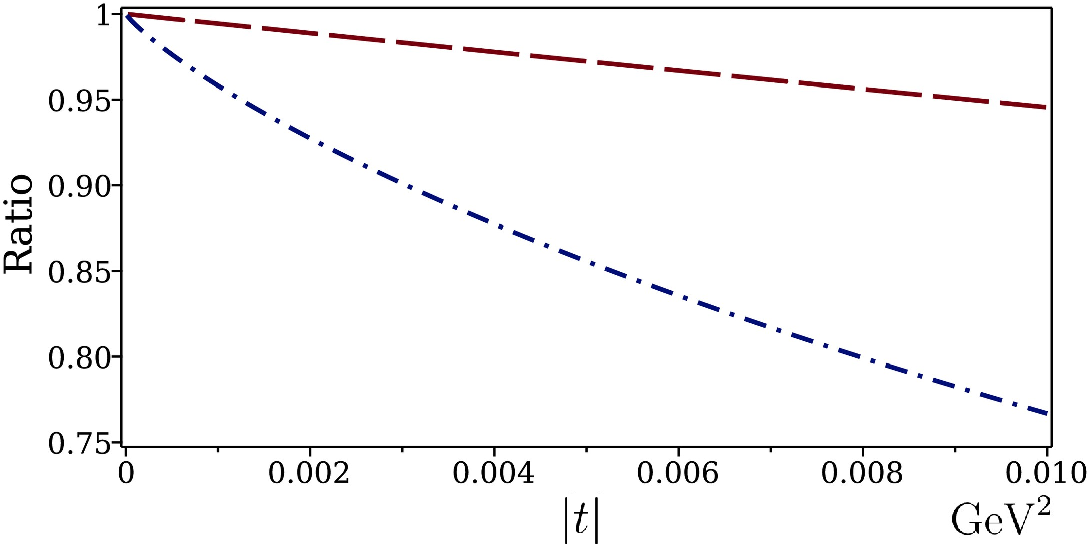}
%\hspace*{0.1\textwidth}
\vspace*{-0.5\baselineskip}
\caption{\small Ratios of the differential cross sections with amplitudes (\ref{A18}) [dashed curve] and (\ref{A19}) [dot-dashed curve] to the differential cross section for scattering of point-like particles.
}\label{Fig1}
\end{center} 
\vspace*{-0.5\baselineskip}
\end{figure*} 

Note that amplitude (\ref{A18}) is linear in the product of form factors. This is fundamentally different from the result (\ref{A14}) in the potential approach with exponential dependence on the form factors. To quantify the difference, we compare the differential cross sections in both approaches. We consider the case of dipole parameterization of form factors with ${\cal F}(q^2) = \left(1+q^2/\Lambda^2\right)^{-2}$, $\Lambda = \sqrt{0.71}$ GeV. Then amplitude (\ref{A14}) takes the form \cite{Petrov2024}
\begin{equation}\label{A19}
F^{C}(s,q^2) e^{-2\\i{\delta}^{C}_{ab}(0)} = - \, \frac{s \alpha}{q^2} \left[ \frac{2}{\Gamma(1 \!-\! \i \alpha)} 
\left(\frac{q}{2\mu}\right)^{1-\\i \alpha} \!
K_{1+\\i \alpha}\left(\frac{q}{\mu}\right)\right] ,
\end{equation}
where $K_{\nu}$ is the McDonald function of the order $\nu$, and $\mu=3\Lambda/10$. The expression in front of the square brackets in (\ref{A19}) represents the amplitude for point-like charges (up to the phase factor), and the expression in square brackets defines the correction due to the finite sizes of the particles. When $q^2 \to 0$ the later expression goes into $\exp\{ \i \alpha \ln (4\mu^2/q^2) + \ln \left[\Gamma(1+\i\alpha) / \Gamma(1-\i \alpha)\right] \}$, which is a pure phase. Fig.~\ref{Fig1} presents the differential cross sections with amplitudes (\ref{A18}) and (\ref{A19}) in relation to the differential cross section of point-like particles. We show region $|t| < 10^{-2} \,\mbox{GeV}^{2}$ relevant to the scattering mode considering here in the case of $p$-$p$ scattering. In this region the Coulomb interaction plays an important role and CNI effect is significant at the LHC energies \cite{Kaspar2011}. At the right end of this region the discrepancy between the cross sections is approximately 19\%. At $|t| = 0.002$ GeV$^2$, where the Coulomb contribution begins to dominate over the strong interaction \cite{TOTEM8,TOTEM13}, the discrepancy is 6\%. At the same time, the statistical error of the cross section measured at the LHC in this region is about 1\%.

So, in the case of proton scattering the difference is significant. This means that it is not reasonable to use the potential approach to describe CNI effect in proton scattering at the LHC. At the same time, in the region $|t| \lesssim 2 \times 10^{-4}\,\mbox{GeV}^{2}$, where the Pb-Pb Coulomb scattering occurs in the same mode, the difference between the two approaches is less than 1\%. This means that the description of Pb-Pb Coulomb scattering in the latter region is practically indistinguishable in both approaches.

\subsection{Composite extended particles \label{sec3.2}}

As the transferred momenta increase from the above values, the internal structure of colliding particles becomes increasingly significant and the mode of multiple scattering on their charged components becomes prevailing. This mode was described by Glauber \cite{Glauber1959} in relation to the elastic scattering due to strong interactions of a fast point-like particle off a nucleus composed of nucleons. However, a generalization to the case of nucleus-nucleus scattering has not yet been found in a compact form. Below we construct the corresponding generalization applicable to Coulomb scattering of composite particles, using the terms of nuclear physics for definiteness. A special case of hadron scattering in this mode, we consider in the next subsection.

So, following the Glauber theory, we assume that protons in colliding nuclei are ``frozen'' as they pass each other, and the impacts of any proton of one nucleus on protons of another nucleus are independent of each other. The former condition means that nuclei are characterized by certain positions of the protons in the transverse projection. The latter condition means the absence of interaction correlations. In this case, the eikonal phase is the sum of individual phase shifts,
\begin{equation}\label{A20} 
\delta_{ab}^C ({\bf b}, \{{\bf s}_{a} , {\bf s}_{b}\}) = 
\sum_{^{i=1, \dots N_a}_{j=1, \dots N_b}} \delta^C ( {\bf b} - {\bf s}_{a_i}\! + {\bf s}_{b_j}) \,.
\end{equation} 
Here ${\bf s}_{a_i}$ and ${\bf s}_{b_j}$ are the positions of the protons $i$ and $j$ in the transverse projection relative to the centers of scattering nuclei $a$ and $b$, respectively, $N_a$ and $N_b$ are the numbers of charged protons inside them, and $\delta^C$ is the elementary Coulomb shift (\ref{A7}). For a given eikonal phase, the scattering amplitude operator acting in the space of variables $\{{\bf s}_{a} , {\bf s}_{b}\}$ is defined as
\begin{equation}\label{A21}
F^{C}(s,q^2;\{{\bf s}_{a},{\bf s}_{b}\}) = 
\frac{s}{4\pi \i} \int \!\d^2 {\bf b} \, 
e^{\\i {\bf q} {\bf b}} \left[ e^{2 \\i \delta_{ab}^C ({\bf b}, \{{\bf s}_{a} , {\bf s}_{b}\})} - 1 \right] .
\end{equation}
The actual amplitude is determined by averaging (\ref{A21}) with the square of the product of wave functions of colliding particles. Practically this means taking integrals $\d^2 \{{\bf s}_{a_i}\!\}$ and $\d^2 \{{\bf s}_{b_j}\!\}$ of (\ref{A21}) with weights defined as the distributions of proton charges that specify the probability distribution of photon emission and absorption. So, the actual amplitude is 
\begin{eqnarray}\label{A22}
&  \displaystyle F^{C}(s,q^2) =  
\int  \d^2 \{{\bf s}_{a}\} \d^2 \{{\bf s}_{b}\} \; 
\varrho_a (\{{\bf s}_{a}\}) \varrho_b (\{{\bf s}_{b}\}) \,
F^{C}(s,q^2;\{{\bf s}_{a},{\bf s}_{b}\}). &
\end{eqnarray}
Here $\d^2 \{{\bf s}_{\kappa}\} = \d^2 {\bf s}_{\kappa_1} \dots \d^2 {\bf s}_{\kappa_{N_\kappa}}$, $\varrho_\kappa (\{{\bf s}_{\kappa}\}) = \varrho_\kappa ({\bf s}_{\kappa_1},\dots{\bf s}_{\kappa_{N_\kappa}})$, $\kappa = a,b$ and $\varrho_\kappa$ is the multi-proton density normalized to unity,
\begin{equation}\label{A23}
\int \d^2 \{{\bf s}_{\kappa}\} \, \varrho_\kappa (\{{\bf s}_{\kappa}\}) =1 \,.
\end{equation} 
Following \cite{Glauber1959}, we further assume that the protons are distributed uncorrelatedly and their spatial densities are factorized:
\begin{equation}\label{A24}                                                                                                                                                                                                                                                                                                                                                    \varrho_\kappa (\{{\bf s}_{\kappa}\}) =  \prod_{i=1, \dots N_{\kappa}} \hat\rho_{\kappa_i}({\bf s}_{\kappa_i}\!)\,,
\end{equation} 
where $\hat\rho_{\kappa_i}({\bf s}_{\kappa_i}\!)$ are the one-proton densities normalized to 1.

In terms of the profile functions of the Coulomb interaction of the protons
\begin{equation}\label{A25}
\gamma^C({\bf b} - {\bf s}_{a_i} + {\bf s}_{b_j}) = 
e^{2\\i \delta^C ( {\bf b} - {\bf s}_{a_i} + {\bf s}_{b_j})} - 1
\end{equation} 
the scattering amplitude operator (\ref{A21}) is written as
\begin{equation}\label{A26}
F^{C}(s,q^2;\{{\bf s}_{a_i} , {\bf s}_{b_j}\}) = 
\frac{s}{4\pi \i} \int \d^2 {\bf b} \; 
e^{\\i {\bf q} {\bf b}} \left\{ \prod_{i=1}^{N_a} \prod_{j=1}^{N_b}
\left[1+\gamma^C({\bf b} - {\bf s}_{a_i} + {\bf s}_{b_j}) \right] - 1 \right\} .
\end{equation}
Then amplitude (\ref{A22}) takes the form
\begin{equation}\label{A27}
F^{C}(s,q^2) = 
\frac{s}{4\pi \i} \int \d^2 {\bf b} \; 
e^{\\i {\bf q} {\bf b}} \; \Gamma^C({\bf b}) \,,
\end{equation}
where $\Gamma^C({\bf b})$ is the Coulomb profile function of scattering nuclei,
\begin{equation}\label{A28}
\!
\Gamma^C({\bf b}) = 
\int  \d^2 \{{\bf s}_{a}\} \d^2 \{{\bf s}_{b}\} \; 
\varrho_a (\{{\bf s}_{a}\}) \varrho_b (\{{\bf s}_{b}\})
\prod_{i=1}^{N_a} \prod_{j=1}^{N_b} 
\left[1+\gamma^C({\bf b} - {\bf s}_{a_i} + {\bf s}_{b_j}) \right] - 1  .
\end{equation}
Opening square brackets in (\ref{A28}), we obtain an expansion in the number of interactions
of the protons. 

Formula (\ref{A28}) describes the general structure of the elastic scattering profile function in the case of multiple Coulomb scattering. A similar formula takes place in the case of nuclear scattering due to strong interactions. Unfortunately, contributions with a number of profile functions greater than 2 are characterized by very complicated combinatorics and are analytically resolved only for the lightest nuclei with a number of nucleons $\le 4$ \cite{Franco}. If the nuclei are heavier, then either the Monier-Carlo method or some approximation must be used.   

For heavy nuclei, the so-called optical approximation was proposed \cite{Czyz,Formanek}, which implies a formal transition to an infinite number of nucleons. In view of this, it is sometimes believed that for a large number of nucleons this approximation defines the leading contribution to the amplitude. This is certainly true if a point-like particle is scattered off a nucleus. However, in the case of scattering of two composite objects on each other a certain set of contributions that do not contain any small parameters is discarded in this approximation \cite{Andreev,Kaidalov}. Below we show that in the case of Coulomb scattering such a small parameter does exist and it is the fine structure constant $\alpha$.

So, following the general methodology of Glauber theory, we assume that all one-proton densities are the same, and introduce the charge density of nucleus containing $N_{\kappa}$ protons, 
\begin{equation}\label{A29}
\rho_\kappa ({\bf s}) = N_{\kappa} \times \hat\rho_{\kappa}({\bf s})\,.
\end{equation} 
So, $\rho_{\kappa}({\bf s})$ is normalized to $N_{\kappa}$, which is the charge number of the nucleus. Given (\ref{A29}) and taking into account (\ref{A24}), we write (\ref{A28}) as 
\begin{equation}\label{A30}
\Gamma^C({\bf b}) = 
\int  \d^2 \{{\bf s}_{b}\} \varrho_b (\{{\bf s}_{b}\}) 
\left[1+\frac{1}{N_a} \int \d^2 {\bf s}_{a} \, 
{\rho_a} ({\bf s}_{a}) \,
\Gamma^C_{N_b}({\bf b} - {\bf s}_{a};\{{\bf s}_{b}\}) 
\right]^{N_a} \!\!- 1  \,.
\end{equation}  
Here ${\bf s}_{a}$ is an integration variable in the product of $N_a$ identical integrals, and  
\begin{equation}\label{A31}
\Gamma^C_{N_b}({\bf b} - {\bf s}_{a};\{{\bf s}_{b}\}) =
\prod_{j=1}^{N_b} [1+\gamma^C({\bf b} - {\bf s}_{a} + {\bf s}_{b_j})] -1\,.
\end{equation} 
If $N_a$ is large enough, then integrand in (\ref{A30}) is well approximated by the exponent, and we arrive at
\begin{equation}\label{A32}
\Gamma^C({\bf b}) = 
\int  \d^2 \{{\bf s}_{b}\} \varrho_b (\{{\bf s}_{b}\}) \,
\exp\!\left[\int \d^2 {\bf s}_{a} \, 
{\rho_a} ({\bf s}_{a}) 
\Gamma^C_{N_b}({\bf b} - {\bf s}_{a};\{{\bf s}_{b}\}) 
\right] - 1  \,.
\end{equation}  
This transformation is called the optical approximation in Glauber theory.\footnote{Available experience shows that it gives an accuracy of about $2 \div 3 \%$ for the number of scattering centers  from 10 or more \cite{Tarasov}.} In the given case, it was done for protons of the nucleus $a$. A remarkable property of (\ref{A32}) is the disappearance of the dependence on $N_a$ and the transition of charge density ${\rho_a} ({\bf s}_{a})$ to the exponent. This can be interpreted in such a way that protons of nucleus $a$ form a kind of continuous medium (cloud) on which the protons of nucleus $b$ are scattered. Accordingly, the expression in the exponent in (\ref{A32}) can be considered as the phase shift of this scattering.

Now we linearize $\Gamma^C_{N_b}({\bf b} - {\bf s}_{a};\{{\bf s}_{b}\})$ in (\ref{A32}) with respect to the fine structure constant $\alpha$, i.e.~retain in the phase shift in (\ref{A32}) only the leading term in its expansion in powers of $\alpha$. Then due to (\ref{A25}) and (\ref{A31}) expression (\ref{A32}) takes the form
\begin{equation}\label{A33}
\Gamma^C({\bf b}) = 
\int  \d^2 \{{\bf s}_{b}\} \varrho_b (\{{\bf s}_{b}\}) \, 
\exp\!\left[ 2\i \!
\int \!\d^2 {\bf s}_{a} \, {\rho_a} ({\bf s}_{a}) 
\sum_{j=1}^{N_b} \delta^C ( {\bf b} - {\bf s}_{a} + {\bf s}_{b_j}) 
\right] - 1  \,.
\end{equation} 
It is worth noting that as a result of the linearization, each proton of nucleus $b$ interacts with the cloud of protons of nucleus $a$ only once.

Formula (\ref{A33}) allows us to perform the optical approximation once again, this time with respect to the protons of nucleus $b$. To do this, we introduce the profile function of scattering of protons of nucleus $b$ on the above-mentioned charged cloud of the nucleus $a$,
\begin{equation}\label{A34}
\bar{\gamma}^C_a ({\bf b} + {\bf s}_{b_j}) = 
\exp\!\left[2\i \! 
\int \!\d^2 {\bf s}_{a} \, {\rho_a} ({\bf s}_{a}) \,
\delta^C ( {\bf b} - {\bf s}_{a} + {\bf s}_{b_j}) 
\right] - 1  \,,
\end{equation} 
and rewrite (\ref{A33}) in the form
\begin{equation}\label{A35}
\Gamma^C({\bf b}) = 
\int  \d^2 \{{\bf s}_{b}\} \varrho_b (\{{\bf s}_{b}\}) \,
\prod_{j=1}^{N_b}
\left[1+\bar{\gamma}^C_a ({\bf b} + {\bf s}_{b_j}) \right] - 1  .
\end{equation} 
Passing to the charge distribution density of nucleus $b$, we arrive at 
\begin{equation}\label{A36}
\Gamma^C({\bf b}) = 
\left[1+\frac{1}{N_b} \int \d^2 {\bf s}_{b} \, 
{\rho_b} ({\bf s}_{b}) \,
\bar{\gamma}^C_a ({\bf b} + {\bf s}_{b}) \right]^{N_b} \!\!- 1  \,.
\end{equation} 
Transiting to the optical approximation in (\ref{A36}) and linearizing $\bar{\gamma}^C_a$ in $\alpha$, we obtain
\begin{equation}\label{A37}
\Gamma^C({\bf b}) = \exp\!
\left[2\i \!
\int \!\d^2 {\bf s}_{a}\d^2 {\bf s}_{b} \; 
\rho_a ({\bf s}_{a}) \rho_b ({\bf s}_{b}) \, 
\delta^C({\bf b} - {\bf s}_{a} + {\bf s}_{b}) \right] - 1  \,.
\end{equation}
This formula defines the leading non-trivial contribution to the Coulomb profile function of heavy nuclei. Note that the symmetry between $a$ and $b$ is restored, both charge density functions are exponentiated, and the dependence on the number of protons in the scattered nuclei disappears.

Based on (\ref{A37}) and (\ref{A27}), we arrive at the final result for the scattering amplitude, which repeats formula (\ref{A13}) but with $\delta^{C}_{ab}(b)$ defined using the densities $\rho_a ({\bf s}_{a})$ and $\rho_b ({\bf s}_{b})$ normalized to $N_a$ and $N_b$, respectively. Since the formulas for the amplitude coincides, the elimination of IR divergences is carried out in the same way as in the previous subsection.

In conclusion, we recall that in the previous subsection formula (\ref{A13}) was obtained in the potential approach in the case of particles as structureless objects with a continuous charge distribution. Here we obtained this formula in the QFT approach in a different kinematic region and in a different scattering mode, when the scattering occurs on a large number of charged components of the particles. The coincidence of the final formula with the previous one occurred after a double transition to the optical approximation and linearization in $\alpha$ of the resulting phase shift. Performing these operations twice leads to an effective scattering pattern as a scattering of charged clouds with a continuous charge distribution. This pattern corresponds to the leading at small $\alpha$ approximation for Coulomb scattering of heavy nuclei.

\subsection{Hadrons as composite extended particles \label{sec3.3}}

The case of elastic Coulomb scattering of hadrons on charged partons is in many ways analogous to the case with nuclei considered above. The differences are due to the fact that the number of partons in fast-moving hadrons is not fixed and they have different electric charges. In this regard, we introduce to the eikonal phase the electric charges $\varepsilon_i$ and $\varepsilon_j$ of the partons in units of $\sqrt{\alpha}$. So instead of (\ref{A20}) we write
\begin{equation}\label{A38} 
\delta_{ab}^C ({\bf b}, \{{\bf s}_{a} , {\bf s}_{b}\}) = 
\sum_{^{i=1, \dots N_a}_{j=1, \dots N_b}} \varepsilon_i \varepsilon_j\, \delta^C ( {\bf b} - {\bf s}_{a_i}\! + {\bf s}_{b_j}) \,.
\end{equation} 
Furthermore, since the number of partons is not fixed, the amplitude is defined as
\begin{equation}\label{A39}
F^{C}(s,q^2) = 
\sum_{\{N_a\},\{N_b\}} W_{\{N_a\}} W_{\{N_b\}} \;
F^{C}_{\{N_a,N_b\}}(s,q^2)\,,
\end{equation} 
\begin{equation}\label{A40}
F^{C}_{\{N_a,N_b\}}(s,q^2) =  
\int  \d^2 \{{\bf s}_{a}\} \d^2 \{{\bf s}_{b}\} \; 
\varrho_a (\{{\bf s}_{a}\}) \varrho_b (\{{\bf s}_{b}\}) \,
F^{C}(s,q^2;\{{\bf s}_{a},{\bf s}_{b}\})\,. 
\end{equation}
Here $\{N_\kappa\}$ is a certain set of $N_\kappa$ charged partons. The probabilities $W_{\{N_\kappa\}}$ satisfy the conditions $\sum_{\{N_\kappa\}} W_{N_{\{N_\kappa\}}} = 1$, where the summation is carried out over the specific sets and the numbers of partons in the sets. 

In accordance with (\ref{A38}), the profile functions of the Coulomb interaction of the partons we now define as,
\begin{equation}\label{A41}
\gamma^C_{ij}({\bf b} - {\bf s}_{a_i} + {\bf s}_{b_j}) = 
e^{2\\i \varepsilon_i \varepsilon_j \delta^C ( {\bf b} - {\bf s}_{a_i} + {\bf s}_{b_j})} - 1\,.
\end{equation} 
Accordingly, the partial amplitude (\ref{A40}) is written as
\begin{equation}\label{A42}
F^{C}_{\{N_a,N_b\}}(s,q^2) = 
\frac{s}{4\pi \i} \int \d^2 {\bf b} \; 
e^{\\i {\bf q} {\bf b}} \; \Gamma^C_{\{N_a,N_b\}}({\bf b}) \,,
\end{equation}
where $\Gamma^C_{\{N_a,N_b\}}({\bf b})$ is the partial profile function of scattering hadrons,
\begin{equation}\label{A43}
\!
\Gamma^C_{\{N_a,N_b\}}({\bf b}) = 
\int  \d^2 \{{\bf s}_{a}\} \d^2 \{{\bf s}_{b}\} \; 
\varrho_a (\{{\bf s}_{a}\}) \varrho_b (\{{\bf s}_{b}\})
\prod_{i=1}^{N_a} \prod_{j=1}^{N_b} 
\left[1+\gamma^C_{ij}({\bf b} - {\bf s}_{a_i} + {\bf s}_{b_j}) \right] - 1  .
\end{equation}

In view of the subsequent averaging over the sets of partons, the combinatorics in (\ref{A43}) must be resolved for arbitrary $N_a$ and $N_b$. However, in the general case this is hardly possible. So, apparently the only way to obtain a workable solution is to use the optical approximation, which entails the disappearance of dependence on $N_a$ and $N_b$. Unfortunately, the condition of a large number of scattering components faces a serious problem in the case of hadrons. 

Really, in a proton at rest the average number of charged partons is 3. In a fast-moving proton their average number is increasing with increasing the energy (due to the increase in the length of parton splitting cascades). In collisions with increasing $q^2$ the effect of parton dissociation is additionally activated. However, in both cases the growth of the average number of partons is very slow (logarithmic). As a result, with restricted momentum transfers the condition for the applicability of the optical approximation in the case of hadrons is extremely high collision energies. Moreover, the energy must be so high that it can hardly be achieved at colliders.\footnote{It is enough to mention that at 100 TeV the multiplication factor of the average number of partons is only 2.15 \cite{Nekrasov2021}.} Nevertheless, if there is an effect, it is worth studying.

So, we proceed to constructing the optical approximation in the case of Coulomb scattering of hadrons. We will carry out the analysis in two stages. At the first stage, as a methodological technique, we assume that the colliding hadrons $a$ and $b$ contain partons of only one type, i.e.~their partons have common charges  and common one-parton densities, although in different hadrons they may be different. In this case the charges of our hypothetical hadrons are $N_{\kappa} \times \varepsilon_{\kappa}$, where $\kappa = a,b$, and their charge distribution densities in units of $\varepsilon_\kappa\sqrt{\alpha}$ are enhanced by $N_{\kappa}$ times the one-parton densities,
\begin{equation}\label{A44}
\widetilde{\rho_\kappa} ({\bf s}) = N_{\kappa} \times \hat\rho_{\kappa}({\bf s})\,.
\end{equation} 
So, $\widetilde{\varrho_\kappa}({\bf s})$ are normalized to $N_{\kappa}$. Here the tilde means that we are considering hypothetical particles. Given (\ref{A44}) and repeating the calculations of the previous subsection, we arrive at the following expression for the partial profile function,
\begin{equation}\label{A45}
\Gamma^C_{\{N_a,N_b\}}({\bf b}) = \exp\!
\left[\int \!\d^2 {\bf s}_{a}\d^2 {\bf s}_{b} \; 
\widetilde{\rho_a} ({\bf s}_{a}) 
\widetilde{\rho_b} ({\bf s}_{b}) \,
2\i \varepsilon_a \varepsilon_b \,
\delta^C({\bf b} - {\bf s}_{a} + {\bf s}_{b}) \right] - 1  \,.
\end{equation}

At the second stage, we return to the case of real hadrons. Namely, we ``allow'' them to consist of partons of certain types and charges. At the same time, we require that the number of partons of each type be large enough. This will allow us to go to the optical approximation separately for each group of partons. In this case, under the exponent sign in the corresponding analogue of formula (\ref{A45}) the sum of contributions of partons of all types will appear. (The transition to the real case can be made first for one particle and then for another.) As a result, $\varepsilon_a \widetilde{\rho_a} ({\bf s}_{a}) \, \varepsilon_b \widetilde{\rho_b} ({\bf s}_{b})$ in (\ref{A45}) will move into the sum of the products of contributions of groups of partons of certain types, and then into the product of charge densities of real hadrons: 
\begin{equation}\label{A46}
\varepsilon_a \widetilde{\rho_a} ({\bf s}_{a}) \, \varepsilon_b \widetilde{\rho_b} ({\bf s}_{b}) \;\longrightarrow \;
\sum_{I} \varepsilon_{a_I} \rho_{a_I} ({\bf s}_a)
\sum_{J} \varepsilon_{b_J} \rho_{b_J} ({\bf s}_{b}) =
{\rho}_{a} ({\bf s}_{a}) {\rho}_{b} ({\bf s}_{b})\,.
\end{equation} 
Here, in the middle expression summation is carried out over the groups of partons, and $\rho_{\kappa_I} ({\bf s}_{\kappa})$ are the charge distribution densities of the groups. In the final expression ${\rho}_{\kappa} ({\bf s}) = \sum_{I} \varepsilon_{\kappa_I} \rho_{\kappa_I} ({\bf s})$, which is the charge density of the real hadron $\kappa$ in units of $\sqrt{\alpha}$.

Now note that due to the absence of dependence on $\{N_a\}$ and $\{N_b\}$ in the final formula for the  partial profile function, which is obtained by combining (\ref{A45}) and (\ref{A46}), the partial amplitude (\ref{A42}) is also independent of $\{N_a\}$ and $\{N_b\}$. Hence, averaging with probabilities $W_{\{N_a\}}$ and $W_{\{N_b\}}$ in (\ref{A39}) is trivial. 

So, we again arrive at formula (\ref{A13}), in this case for the amplitude of Coulomb scattering of hadrons as structured composite particles. The $\delta^{C}_{ab}(b)$ in this formula is defined with the charge densities of real hadrons, as defined above. Our concluding remarks repeat the last paragraph of the previous subsection, with ``heavy nuclei'' replaced by ``hadrons at extremely high energies''.

\section{Discussion and conclusion \label{sec4} }

We have carried out a description in the eikonal model of high-energy Coulomb elastic scattering of extended particles in three scattering modes determined by the magnitude of the momentum transfer, and we have compared the predictions of the model in the potential and QFT approaches. 

In the limit of small transfers, particles are scattered as point-like objects, and the description of their Coulomb scattering coincides in both approaches. In essence, this is a well-known result, and we present it for completeness. However, if the wavelength of the exchanged photons becomes comparable to the particle size, a significant difference appears between the potential and QFT approaches. In this case the scattering occurs between the volumes of colliding particles, but the methods for taking into account their form factors in both approaches are different. The upper boundary of the momentum transfer for establishing this scattering mode can be estimated as a value inverse to the sum of transverse radii of the colliding particles. In the case of proton scattering this gives $q^2 \lesssim 10^{-2} \,\mbox{GeV}^{2}$. (In the same region CNI effect is significant.) In the case of heavy nuclei scattering, using the example of Pb-Pb scattering, a similar estimate is $q^2 \lesssim 2\times 10^{-4}\,\mbox{GeV}^{2}$.

The reason for the difference is related to the relativistic non-invariance of the Schr{\"o}dinger equation which underlies the potential approach. In particular the expression for the kinetic energy of particles of the type $\vec{\, p}^{\,2}/2m$ is incorrect in the relativistic region. However, this expression is used significantly in the derivation of formula (\ref{A3}) for the eikonal shift. At the same time, the relativistic non-invariance of the potential appearing in this formula is irrelevant. In classical electrodynamics it transforms in a covariant manner, and this effect together with the  potential retardation effect can be taken into account. Both of them do not affect the eikonal shift due to the integration of the potential in formula (\ref{A3}) over the longitudinal coordinate. In quantitative terms, the difference between the descriptions in the two approaches is significant in the CNI region in the case of proton scattering. We interpret this as an inappropriateness of the potential approach to the description of the proton Coulomb scattering at the LHC energies. At the same time, in the case of heavy nuclei Coulomb scattering in the relevant momentum transfer region, the quantitative discrepancy between the descriptions in the two approaches is insignificant.

As $q^2$ increases from the above values, the Coulomb scattering switches to the mode with scattering on individual charged components of colliding particles. As a result, the scattering takes on a multiple character, which is described by Glauber theory. In the case of nuclei this mode becomes dominant at $q^2 \gtrsim 10^{-2} \,\mbox{GeV}^{2}$, and the mentioned components are initially charged protons and with a further increase in $q^2$ they become charged partons in the nucleons. If nuclei contain a large number of protons, then the scattering amplitude is described by formula (\ref{A13}) with the modification indicated in the end of Section \ref{sec3.2}. The fundamentally new result here is that the optical approximation underlying this formula is indeed the leading one at $\alpha \to 0$ in the case of heavy nuclei. Additionally, it can be noted that formula (\ref{A13}) can apparently be used as a first approximation in the entire region of the momentum transfer allowed in the eikonal approach, since in the region of scattering of nuclei as structureless objects the same formula describes the amplitude acceptably (see above discussion). 

In hadron collisions, the situation with increasing $q^2$ beyond the limit specified above is more complicated, since in this case an analogue of a large number of protons in nuclei is a large number of charged partons in hadrons. For restricted $q^2$ this can be achieved at extremely high collision energies, certainly above 100 TeV, since the growth with the energy of the average number of partons is very slow (logarithmic). Unfortunately, at the energies actually achievable at colliders, the Coulomb scattering of hadrons falls into an intermediate zone, in which a compact solution has not been found. Fortunately, the solution in this region has no practical significance, since at $q^2 > 10^{-2} \,\mbox{GeV}^{2}$ the Coulomb interaction of hadrons becomes negligible against the background of their strong interactions. Nevertheless, the theoretical description of this scattering mode is of interest because it sheds light on the conditions under which the potential scattering formula, used in most studies within the eikonal approach, may become relevant for describing Coulomb scattering of hadrons.

In conclusion, we recall that Coulomb contributions become dominant in the region of very small momentum transfer in elastic scattering of strongly interacting charged particles. In this regard, the description of Coulomb scattering is an essential component of the study of their interactions at modern colliders. Meanwhile, until now there was no understanding of how to correctly take into account the finite sizes of colliding particles when describing their Coulomb interactions in the eikonal model. In this paper, we have shown that in the case of proton collisions at the LHC, the inclusion of form factors directly in the eikonal phase is erroneous and leads to significant inaccuracies. The correct way is to consider them as factors in full amplitude. There was also no understanding whether the use of the optical approximation in describing  Coulomb scattering of heavy nuclei was correct. We have shown that this approximation is leading in the limit of smallness of the fine structure constant~$\alpha$.

\bigskip

\noindent {\it Acknowledgments}: The author is grateful to V.A.~Petrov for stimulating discussions.

\end{document}